		\newif\ifpreprint%
		\preprinttrue
		\newif\iffulldoc
		\fulldoctrue
\newcommand{\commonoptslist}%
	{%
	aps,%
	prd,%
	showpacs,%
	floatfix,%
	amsmath,%
	amssymb,%
	amsfonts,%
	nofootinbib,%
	superscriptaddress,%
	}%
\ifpreprint
\documentclass[preprint,\commonoptslist]{revtex4}
\else
\documentclass[twocolumn,\commonoptslist{}]{revtex4}
\fi
			\usepackage{graphicx}		
			\usepackage{ifthen}		
			\usepackage{natbib}		
			\usepackage{abbrvate}		
			\usepackage{mathrsfs}		
			\usepackage{bbm}			
			\usepackage{wasysym}
			\newcommand{\imgpath}{Images}
			\newcommand{\bibpath}{bibtex}
			
			\newabbr{AMSB}{anomaly mediated supersymmetry breaking}
			\newabbr{BSM}{beyond the standard model}
			\newabbr{CISLR}{conformal invariant supersymmetric left-right}
			\newabbr{CKM}{Cabbibo-Kobayashi-Maskawa}
			\newabbr{EAM}{extended anomaly mediation}
			\newabbr{EWSB}{electroweak symmetry breaking}
			\newabbr{GAYMS}{gauge-anomaly-yukawa mediated supersymmetry breaking}
			\newabbr{GMSB}{gauge mediated supersymmetry breaking}
			\newabbr{GUT}{grand unified theory}
			\newabbr{IR}{infrared}
			\newabbr{LEP}{large electron positron collider}
			\newabbr{LSP}{lightest supersymmetric particle}
			\newabbr{LHC}{large hadron collider}
			\newabbr{mAMSB}{minimal anomaly mediated supersymmetry breaking}
			\newabbr{mGMSB}{minimal gauge mediated supersymmetry breaking}
			\newabbr{MSSM}{minimal supersymmetric standard model}
			\newabbr{mSUGRA}{minimal supergravity}
			\newabbr{NLSP}{next-to lightest supersymmetric particle}
			\newabbr{NMSSM}{next-to minimal supersymmetric standard model}
			\newabbr{RGE}{renormalization group equation}
			\newabbr{SM}{standard model}
			\newabbr{SSB}{spontaneous symmetry breaking}
			\newabbr{SUGRA}{supergravity}
			\newabbr{SUSY}{supersymmetry}
			\newabbr{SUSYLR}{supersymmetric left-right}
			\newabbr{UV}{ultra violet}
			\newabbr{VEV}{vacuum expectation value}
				\newcommand{\eat}[1]{}
				\newcommand{\Ybar}{\overline{Y}}

\newcommand{\eq}[1]{Eq.~\eqref{Eq:#1}}
\newcommand{\eqn}[1]{\eqref{Eq:#1}}
\newcommand{\fig}[1]{Figure~\ref{Fig:#1}}

\newcommand{\Sec}[1]{Section~\ref{Sec:#1}}

			\newcommand{\Fphi}{F_{\phi}}
			
			\newcommand{\Mmess}{M_\text{mess}}
			
		\newcommand{\pderiv}[3][]{%
			\ifthenelse{\equal{#1}{}}%
				{%
				\frac{ \partial #2}{ \partial #3 }%
				}%
				{%
				\frac{ \partial^{#1} #2}{ \partial {#3}^{#1} }%
				}%
			}
		\newcommand{\deriv}[3][]{%
			\ifthenelse{\equal{#1}{}}%
				{%
				\frac{ d #2}{ d #3 }%
				}%
				{%
				\frac{ d^{#1} #2}{ d {#3}^{#1} }%
				}%
			}
		\newcommand{\fderiv}[3][]{%
			\ifthenelse{\equal{#1}{}}%
				{%
				\frac{ \delta #2}{ \delta {#3} }%
				}%
				{%
				\frac{ \delta^{#1} #2}{ \delta {#3}^{#1} }%
				}%
			}
		
		\newcommand{\intOp}[2][]{\int \! d^{#1}#2 \;}
		\newcommand{\abs}[1]{ \mathopen{}\left| {#1}\right| }

		\newcommand{\commutator}[2]{\inb{#1,#2}}

		\newcommand{\half}{\frac{1}{2}}			
		\newcommand{\fourth}{\frac{1}{4}}		
		\newcommand{\eighth}{\frac{1}{8}}		

		\newcommand{\inp}[2][0cm]{\mathopen{}\left(#2\parbox[h][#1]{0cm}{}\right)}
		\newcommand{\inb}[2][0cm]{\mathopen{}\left[#2\parbox[h][#1]{0cm}{}\right]}
		\newcommand{\inbr}[2][0cm]{\mathopen{}\left\{#2\parbox[h][#1]{0cm}{}\right\}}
		\newcommand{\inap}[2][0cm]{\mathopen{}\left<#2\parbox[h][#1]{0cm}{}\right>}
		
	\newcommand{\vev}[1]{\inap{#1}}

	\newcommand{\superf}[1]{\mathcal{#1}}
	
	\newboolean{ulinescalar}
	\newcommand{\scalarunderline}{\setboolean{ulinescalar}{true}}%
	\newcommand{\scalar}[1]{%
		\ifthenelse{ \boolean{ulinescalar} }
			{
			\ifx#1\DeltaC%
				{%
				\underline{\Delta}^c%
				}%
			\else\ifx#1\DeltaBarC%
				{%
				\underline{\bar{\Delta}}^c%
				}%
			\else\ifx#1\Phi%
				{%
				\underline{\Phi}%
				}%
			\else\ifthenelse{ \equal{#1}{L^c}}%
				{%
				\underline{L}^c%
				}%
			{\ifthenelse{ \equal{#1}{Q^c}}%
				{%
				\underline{Q}^c%
				}%
			{\ifthenelse{ \equal{#1}{H_u}}%
				{%
				\underline{H}_u%
				}%
			{\ifthenelse{ \equal{#1}{H_d}}%
				{%
				\underline{H}_d%
				}%
				{%
				\underline{#1}%
				}}}}%
			\fi\fi\fi
			}
			{
			#1%
			}
		}%

\begin{document}

\abbrstyle{expandall}
\begin{abstract}
Despite its successes---such as solving the supersymmetric flavor problem---\abbr{AMSB} is untenable because of its prediction of tachyonic sleptons.  An appealing solution to this problem was proposed by Pomarol and Rattazzi where a threshold controlled by a light field deflects the \abbr{AMSB} trajectory, thus evading tachyonic sleptons.   In this paper we 
examine an alternate class of deflection models where the non-supersymmetric threshold is accompanied by a heavy, instead of light, singlet.  
The low energy form of this model is the so-called extended anomaly mediation proposed by Nelson and Weiner, but with potential for a much higher deflection threshold.
The existence of this high deflection threshold implies that 
the space of \abbr{AMSB} deflecting models is larger than previously thought.
\end{abstract}

	\title{When Anomaly Mediation is UV Sensitive}
	\author{N. Setzer}
	\email{nsetzer@unimelb.edu.au}
	\affiliation{%
			School of Physics,
			University of Melbourne,
			Victoria, 3010,
			Australia%
		}
	\author{S. Spinner}
	\email{sspinner@wisc.edu}
	\affiliation{%
			Department of Physics,
			University of Wisconsin,
			Madison, WI 53706,
			USA%
		}
	\date{\today}
	\pagestyle{plain}	
	\scalarunderline
	\abbrstyle{plain}
	\abbrreset
	\abbrmakeused[CKM]
	\abbrmakeused[LEP]
	\abbrmakeused[LHC]

\maketitle

\section{Introduction}

A puzzling aspect of the \abbr{SM} is the stability of the Higgs mass at the electroweak scale---a mystery which is elegantly resolved by \abbr{SUSY}.  Of course a realistic theory requires \abbr{SUSY} to be broken and constructing viable models for \abbr{SUSY} breaking is not a trivial issue.  The challenge, known as the \abbr{SUSY} flavor problem, is to find a scenario which inherently suppresses the contributions to flavor and $CP$ violating operators---while hopefully also significantly reducing the number of \abbr{SUSY} breaking parameters.

One such scenario, particularly successful in terms of the \abbr{SUSY} flavor problem, is \abbr{AMSB}\cite{Randall:1998uk, Giudice:1998xp} (for a review see \cite{Luty:2005sn, Shirman:2009mt}).  To realize \abbr{AMSB} requires a true sequestering of the hidden and visible sectors so that only gravitational interactions exist between the two.  This was originally achieved via a specific geometry in a model with extra dimensions \cite{Randall:1998uk} and then later shown to be possible in four dimensions with a specific type of a superconformally invariant \abbr{SUSY} breaking sector \cite{Luty:2001jh, Luty:2001zv}.  The \abbr{SUSY} breaking is thus intimately linked with the breaking of superconformal invariance, broken at loop level.  The soft \abbr{SUSY} breaking terms are related to a mass scale $\Fphi$ (the gravitino mass), the low energy beta-functions, and the anomalous dimensions.  They are also \abbr{RGE} scale invariant, \abbr{UV} insensitive, highly predictive, and absent of new flavor violation.  

Regardless of these features, \abbr{AMSB} is problematic since naively applying it to the \abbr{MSSM} produces tachyonic slepton masses.  The lack of free parameters and the insensitivity of the soft parameters to \abbr{UV} physics makes this problem challenging.  Still solutions exist: adding low energy Yukawa couplings to alter the beta functions and anomalous dimensions\cite{Chacko:1999am, Allanach:2000gu, Mohapatra:2008gz}, utilizing $D$-term contributions to soft breaking masses\cite{Jack:2000cd, Arkani-Hamed:2000xj, Carena:2000ad}, considering low energy threshold effects \cite{Katz:1999uw}, and deflecting from the \abbr{AMSB} trajectory \cite{Pomarol:1999ie} (with a variety of models along this theme \cite{Kitazawa:2000ft, Okada:2002mv, Nelson:2002sa, Hsieh:2006ig, Kikuchi:2008gj, Everett:2008qy, Everett:2008ey, Altunkaynak:2010xe}).

In the deflection models, a superconformal violating threshold introduces new \abbr{SUSY} breaking causing a deflection from the anomaly mediated trajectory.  Typically this is achieved via messengers that are charged under the \abbr{SM} resulting in mixed anomaly and gauge mediated supersymmetry breaking (for reviews of \abbr{GMSB} see for example \cite{Giudice:1998bp, Martin:1997ns}).  The deflecting threshold in \cite{Pomarol:1999ie} is achieved through a shallow potential yielding a high messenger scale (much larger than $\Fphi$) and a light singlet (modulus); therefore, such deflection is sometimes called light singlet deflection.

Here we offer a scenario (which we shall call heavy singlet deflection) that contains a high-scale \abbr{SUSY} threshold which, in the \abbr{SUSY} limit, has no effect on the \abbr{AMSB} trajectories.  Turning on \abbr{SUSY} breaking induces a second threshold which is \abbr{SUSY} violating and leads to deflection at that scale.  The deflection terms induced are precisely of the \abbr{EAM} form\cite{Nelson:2002sa, Hsieh:2006ig} 
making the theory a \abbr{UV} completion of \abbr{EAM}; however, the induced threshold need not be near $F_\phi$ (as is assumed for \abbr{EAM}) and indeed can generically be much higher.  As such, we will demonstrate that deflection can occur at thresholds well above $F_\phi$ despite the absence of a light-singlet.  The existence of this arbitrarily high deflection threshold indicates that the space of \abbr{AMSB} deflecting models is larger than hitherto expected and that \abbr{AMSB} is not as \abbr{UV} insensitive as commonly believed.

One of the main points of this paper is to clarify that deflection, in general, arises whenever a threshold is controlled by a \abbr{VEV} stabilized by a \abbr{SUSY} breaking term.  These terms break the conformal invariance of \abbr{AMSB} and convey that information through the threshold.  Ref.~\cite{Pomarol:1999ie} proposed a scenario where this deflection included a light singlet; however, a light singlet is not a requirement of a high-scale deflecting threshold (as the existence of heavy singlet deflection demonstrates).  To demonstrate these points on deflection, we employ a simple example often quoted in the literature as a model without deflection, but which actually contains deflection of the heavy singlet type.



Following a review of \abbr{AMSB} and the necessary conditions for deflection in \Sec{AMSB}, this paper presents, in \Sec{ex.model}, an example model illustrating the ideas discussed above.  The latter portion of \Sec{ex.model} is devoted to the discussion of a model employing this mechanism to generate non-tachyonic slepton masses, which resembles the first model presented in \cite{Nelson:2002sa}.

\section{\abbr[<<]{AMSB} and Decoupling}
\label{Sec:AMSB}

In \abbr{AMSB}, only gravitational interactions link the \abbr{SUSY} breaking sector to visible sector and therefore gravity communicates the \abbr{SUSY} breaking.  A convenient way to describe this transmission of \abbr{SUSY} breaking is to separate out an auxiliary component of the gravity supermultiplet called the conformal compensator, $\phi$.  Superconformal invariance then dictates how $\phi$ appears in the lagrangian, and using canonically normalized fields one factor of $\phi$ appears per unit dimension of the coupling; that is $M \to \phi M$.

Due to the \abbr{SUSY} breaking in the sequestered sector, the conformal compensator picks up a non-zero $F$ component; this may be parameterized as
\begin{equation}
\phi = 1 + \Fphi \theta^2.
\label{Eq:phi.components}
\end{equation}
With this choice \abbr{AMSB} predicts the form of the \abbr{SUSY} breaking terms\footnote{Our convention is that $\deriv{\ln Z_i^j}{\ln \mu} = \gamma_i^j$ so that, for example, a yukawa coupling in the superpotential $W \supset \frac{1}{3!} Y^{ijk} \Phi_i \Phi_j \Phi_k$ has a beta function of $\beta_{Y}^{ijk} = - \half Y^{ijp} \gamma_p{}^k + \inp{i \leftrightarrow k} + \inp{j \leftrightarrow k}$\cite{Martin:1993zk}.  We also choose all \abbr{SUSY} breaking terms to come with a plus sign in the potential.}:
\begin{align}
\inp{m^2}_i{}^j
	& =	- \eighth \abs{\Fphi}^2 \commutator{\gamma^\dagger}{\gamma}_i{}^j
		- \fourth \abs{\Fphi}^2 \inb{ \half \pderiv{\gamma_i{}^j}{g_G} \beta_{g_G} + \pderiv{\gamma_i{}^j}{Y^{\ell m n}} \beta_Y^{\ell m n} + \text{h.c.}}
\label{Eq:AMSB.scalar.masses}
	\\
a^{ijk}
	& =	  \Fphi \beta_Y^{ijk}
\label{Eq:AMSB.trilinear.a}
	\\
M_{G}	& =	- \frac{\beta_{g_G}}{g_G} \Fphi,
\label{Eq:AMSB.gaugino.masses}
\end{align}
which may be derived using \eq{phi.components} and promoting both the wave function renormalization constant, $Z_i{}^j$, and inverse gauge coupling, $1/\alpha_G$, to the superfields $\superf{Z}_i{}^j$, $\tau_G$\cite{Randall:1998uk}

The \abbr{AMSB} expressions of Eqs.~\eqn{AMSB.scalar.masses}--\eqn{AMSB.gaugino.masses} are not just boundary conditions; rather, they also represent solutions to the \abbr[>-0+s]{RGE}s.  This unique property is a result of their being determined by superconformal invariance and their origin from quantum corrections.  Since these equations solve the \abbr[>-0+s]{RGE}s, they have been labeled \abbr{AMSB} trajectories---that is, they form a set of equations that may be evaluated at any renormalization scale.

In addition to being solutions of the \abbr{RGE}s, the \abbr{AMSB} trajectories are \abbr{UV} insensitive; in other words, they generically decouple heavy thresholds.  One method to see why this decoupling occurs is to consider what additional \abbr{SUSY} breaking results from the introduction of a threshold---if it is not comparable in size to the \abbr{AMSB} contribution then it may be safely neglected.  

To obtain what new \abbr{SUSY} breaking might be generated by introduction of an intermediate threshold, consider a mass scale $M$ with $\Lambda \gg M \gg F_\phi$ and $\Lambda$ some cutoff for new physics.  Additionally, split the set of superfields $\inbr{\Phi_i}$ into two sets: $q = \{\Phi_i | M_{\Phi_i} \sim M\}$ and $Q = \{\Phi_i | M_{\Phi_i} \ll M\}$, so that $q$ are heavy fields integrated out at the threshold $M$ and $Q$ are light fields remaining in the theory below $M$.  

The most generic lagrangian above $M$ is
\begin{equation}
\mathcal{L}^+ = \mathcal{L}_Q^+ + \mathcal{L}_{Q q} + \mathcal{L}_q
\label{Eq:AMSB.lagrangian.general.threshold.above}
\end{equation}
where $\mathcal{L}_Q^+$ involves only the light fields, $\mathcal{L}_q$ involves only the heavy fields, and $\mathcal{L}_{Qq}$ contains all interactions between the two.  Integrating out the heavy fields yields
\begin{equation}
\mathcal{L}^-
	=	\mathcal{L}_Q^- + M^4 + \ln\inp{\frac{\mu}{M}} f_0\inp{Q} + \frac{1}{M} f_1\inp{Q} + \frac{1}{M^2} f_2\inp{Q} + \dotsb
\label{Eq:AMSB.lagrangian.general.threshold.below}
\end{equation}
which, apart from the cosmological constant term, has at most logarithmic dependence on $M$ (which is induced by quantum corrections).  Absorbing the log dependence into $\mathcal{L}_Q^-$ (since it belongs to the wave function renormalization anyway), the lagrangian below the threshold may be recast schematically as
\begin{multline}
\mathcal{L}^-
	=	  \mathcal{L}_Q^- 
		+ M^4 
		+ \inb[1.25cm]{ \intOp[2]{\theta}
			\inb{
				  \frac{Q^4}{M \phi}
				+ \frac{Q^5}{M^2 \phi^2}
				+ \cdots
			}
			+ \text{h.c.}
			}
		\\ {}
		+ \intOp[4]{\theta}
			\inb{
				  \frac{\inp{Q^\dagger Q}^2 }{M^2 \phi^\dagger \phi}
				+ \frac{\inp{Q^\dagger Q}^3 }{M^4 \inp{\phi^\dagger \phi}^2}
				+ \cdots
			}
\label{Eq:AMSB.lagrangian.schematic.threshold.below}
\end{multline}

The additional \abbr{SUSY} breaking introduced by the threshold can now be evaluated.  Assuming the threshold does not introduce any relevant\footnote{in the renormalization sense of the word} operators---that is, $\mathcal{L}_Q^+ = \mathcal{L}_Q^-$---the portion involving only $Q$'s must, by assumption, respect the superconformal invariance.  Thus, the only \abbr{SUSY} breaking terms that did not exist above $M$ are\footnote{We adopt the notation that an underline indicates the scalar component of a superfield.}
\begin{multline}
\Delta\mathcal{L}^-_{\text{SB}}
	=	\inb{
			  \frac{F_Q}{M} \scalar{Q}^3
			- \frac{F_\phi}{M} \scalar{Q}^4
			+ \frac{F_Q}{M^2} \scalar{Q}^4
			+ \text{h.c.}
			}
		\\
		+ \frac{\abs{F_Q}}{M^2} \abs{F_Q} \abs{\scalar{Q}}^2
		- \frac{F_Q^*}{M^2} F_\phi \scalar{Q} \abs{\scalar{Q}}^2
		- \frac{F_Q}{M^2} F_\phi^* \scalar{Q}^* \abs{\scalar{Q}}^2
		+ \frac{\abs{F_\phi}^2}{M^2} \abs{\scalar{Q}}^4
		+ \cdots
\label{Eq:AMSB.lagrangian.schematic.threshold.below.new.susy.breaking}
\end{multline}
which may be read off of \eq{AMSB.lagrangian.schematic.threshold.below}.  

The salient issue is to then determine their size compared to the \abbr{AMSB} contribution.  By assumption, $F_\phi \ll M$, $\vev{\scalar{Q}} \ll M$, with the latter condition following from the requirement that $Q$ is a light field (that is, if $\vev{\scalar{Q}} \sim M$, it would obtain a mass $M$ and be a heavy field).  Due to these considerations, the only terms that can potentially contribute \abbr{SUSY} breaking comparable to those of \abbr{AMSB} are those that involve the ratio $F_Q/M$.  If it is assumed that $\vev{F_Q} \ll M F_\phi$, then all the terms of \eq{AMSB.lagrangian.schematic.threshold.below.new.susy.breaking} are much smaller than the \abbr{SUSY} breaking contributions due to $\Fphi$ so they may indeed be safely neglected.

The argument is then complete: assuming $\mathcal{L}_Q^+ = \mathcal{L}_Q^-$ and $F_Q \ll M \Fphi$ implies that no significant \abbr{SUSY} breaking can result from the threshold $M$ and the only important contribution is due to \abbr{AMSB}.  Since the threshold does not provide any significant \abbr{SUSY} breaking, the \abbr{AMSB} expressions must be valid, and as they are correct at any scale, they can just be evaluated at scales below $M$.  This result---that the \abbr{AMSB} form may be employed below $M$ whilst maintaining ignorance of the theory above $M$---is the previously mentioned \abbr{UV} insensitivity of \abbr{AMSB}.

\subsection{Pomarol and Rattazzi's Light Singlet Deflection}

Given that \abbr{AMSB}'s \abbr{UV} insensitivity is dependent on two assumptions, it is worthwhile to look for any scenarios that may violate these conditions.  Since the \abbr{AMSB} expressions are a result of superconformal invariance, these violations must arise from breaking this symmetry; that is, they may only result from fields which acquire a \abbr{VEV} when $\Fphi \ne 0$.  A subset of this condition was first considered by Pomarol and Rattazzi\cite{Pomarol:1999ie}, where they discovered that remnant light singlets can have large $F$-term \abbr[>-0+s]{VEV}s.  The argument proceeds as follows: consider the general form for the \abbr{VEV} of the auxiliary component
\begin{equation}
\vev{F_Q} = L_Q + M_Q \vev{\scalar{Q}} + \frac{1}{2!} Y_Q \vev{\scalar{Q}}^2 + \dotsb.
\end{equation}
Since $Q$ is a light field, $\vev{\scalar{Q}} \ll M$ so that neither the Yukawa couplings or any higher dimensional operators can yield a \abbr{VEV} of the desired size.  This leaves the mass term and the linear term; however, the mass term can not generate an $F_Q$ \abbr{VEV} of the needed size: while it is possible $\vev{\scalar{Q}} = \Fphi \ll M$, the condition of $Q$ being light means $M_Q \ll M$ so that $M_Q \vev{\scalar{Q}}$ is necessarily much smaller than $M \Fphi$.  Thus the only term that can yield a large enough $F$ \abbr{VEV} is the linear term $L_Q$.  Since any field charged under the residual gauge symmetry below $M$ would be forbidden from having a linear term, the field whose auxiliary \abbr{VEV} is large must be a singlet of the surviving gauge group.

It is then natural to ask how light this singlet---which shall be called $S$---must be; that is, is it possible to raise the mass of $S$ to the threshold $M$?  To answer this question, assume that the field $S$ has a \abbr[>-1+ic]{SUSY} mass $\mu$.  The lagrangian is\cite{Pomarol:1999ie}
\begin{equation}
\mathcal{L}_S = \intOp[4]{\theta} \inb[7.5mm]{ S^\dagger S + c M \phi^\dagger S + c M \phi S^\dagger } + \inb{ \intOp[2]{\theta} \half \mu \phi S^2 + \text{h.c.} }
\end{equation}
with $c$ some (real) order one constant.  The equations of motion for the auxiliary component gives
\begin{equation}
\vev{F_S} = - c M \Fphi - \mu \vev{\scalar{S}}
\label{Eq:pomarolandrattazzi.vev.singlet.Fterm}
\end{equation}
so that the $F$-term is indeed of the correct size. Varying the lagrangian with respect to $\scalar{S}$ yields the constraint
\begin{equation}
\mu \inp[5mm]{ \vev{F_S} + \Fphi \vev{\scalar{S}} } = 0
\label{Eq:pomarolandrattazzi.vev.singlet}
\end{equation}
which implies that either $\mu = 0$ or $\vev{F_S} = - \Fphi \vev{S}$; however, the latter condition cannot be satisfied since $\vev{\scalar{S}} \ll M$.  A consistent system then demands we take $\mu = 0$.  As the singlet is not allowed a \abbr[>-1+ic]{SUSY} mass term, it must get its mass from the \abbr{SUSY} breaking, so that the singlet's mass is at or below $\Fphi$.  

The result, as argued by Pomarol and Rattazzi\cite{Pomarol:1999ie}, is that having light singlets in the low-scale theory permits the threshold to introduce \abbr{SUSY} breaking effects comparable to $\Fphi$ yielding deflection.  As Pomarol and Rattazzi's scheme requires singlets with mass around $\Fphi$, this particular version is called light singlet deflection.

\subsection{Heavy Singlet Deflection}

An alternative deflection scenario would arise if the threshold introduces relevant operators into $\mathcal{L}_Q^-$ which contain a non-superconformal $\phi$ coupling.  Such a coupling can not arise from any explicit \abbr{SUSY} mass term in $\mathcal{L}_Q^+$ because superconformal invariance is respected by the \abbr{SUSY} preserving lagrangian.  Therefore, the most promising origin of a superconformal violating term would be the \abbr{VEV} of some field which depended on $\Fphi$ (as the \abbr{SUSY} breaking sector does not respect superconformal invariance).  Assuming that the $Q$ fields are all charged under the surviving gauge group (to avoid light singlet deflection), as well as insisting this field not break the residual gauge symmetry, means that no $Q$ field can be responsible for this term.  Furthermore, as the $Q$s must come in at least pairs (to retain gauge invariance), the field whose \abbr{VEV} imparts a superconformal breaking term in $\mathcal{L}_Q$ would have, at best, a yukawa coupling to the $Q$s.  This implies that when said field acquires a \abbr{VEV}, it generates a mass term for the $Q$ fields.  This is important because it means that the \abbr{VEV} can not be of order $M$ (as this would result in the $Q$ fields being heavy), and so whatever field produces the superconformal violation in $\mathcal{L}_Q$ can not be responsible for the threshold $M$.  

While the superconformal violating \abbr{VEV} can not be responsible for the threshold $M$, it is possible that the threshold generates the superconformal violating term.  The basic idea would be that when the superconformal preserving threshold $M$ is created (however that occurs), it induces a small superconformal violating \abbr{VEV} in one of the other $q$ fields, say $q^\prime$.  After the theory is written in the true vacuum (where $\vev{q^\prime} = 0$), any field $Q$ having a yukawa coupling to $q^\prime$ would have a corresponding superconformal violating coupling in $\mathcal{L}_Q^-$ from $q^\prime \rightarrow \vev{q^\prime}$.  The resulting theory below $M$ would not have any light singlets, but would have deflection from the \abbr{AMSB} trajectories.  The next section will discuss an example model to make this idea more concrete.

\section{An Example Model}
\label{Sec:ex.model}

Surprisingly, a model commonly used to demonstrate \abbr{AMSB} decoupling\cite{Pomarol:1999ie,Katz:1999uw} also generates non-superconformal couplings in the low-scale theory.  The superpotential is
\begin{equation}
\superf{W}_0
	= \inp{ \lambda_X X^2 - M^2 \phi^2 } S,
\label{Eq:Toy.SuperW}
\end{equation}
where $X$ and $S$ are singlets and an $R$ symmetry is assumed ($S \to -S$, all other fields invariant) to forbid an explicit mass term or linear term for $X$.  A cubic term for $S$ as well as a mass-mixing term for $X$ and $S$ are still allowed; however, they do not alter the discussion and are therefore omitted for simplicity.  The resulting lagrangian for \eq{Toy.SuperW} is\footnote{We again remind the reader that an underline indicates the scalar component of a superfield.}
\begin{equation}
\begin{aligned}[b]
\mathcal{L}
	&\supset
		  \half \abs{F_X}^2 + \half \abs{F_S}^2
		+ \inp{ 2 \lambda_X \scalar{X} F_{X} - 2 M^2 \Fphi } \scalar{S}
		+ \inp{ \lambda_X \scalar{X}^2 - M^2 } F_S
		+ \text{h.c.}
\end{aligned}
\label{Eq:Toy.L}
\end{equation}
By solving the equations of motion for the auxiliary fields, the $F$-terms
\begin{align}
-F_{S}^*	& =	  \lambda_X \scalar{X}^2 - M^2 & 
-F_X^*	& = 2 \lambda_X \scalar{X} \, \scalar{S}
\label{Eq:Toy.F.term}
\end{align}
are obtained.  In the \abbr[>-1+ic]{SUSY} limit $\vev{F_S} = \vev{F_X} = 0$ so that
\begin{align}
\vev{\scalar{X}}		& = \frac{M}{\sqrt{\lambda_X}}			&
\vev{\scalar{S}}		& = 0
\label{Eq:Toy.VEV.SUSY}
\end{align}
After $X$ obtains a \abbr{VEV}, the superfields $X$ and $S$ mix through a \abbr[>-1+ic]{SUSY} mass term,
\begin{equation}
\superf{W} \supset 2 \sqrt{\lambda_X} M X S,
\label{Eq:Toy.SUSY.mass.X.S}
\end{equation}
leaving both fields heavy so that light singlet deflection does not occur.  This well-established result implies that the superfield $X$ has a \abbr{VEV} given by $\vev{X} = \vev{\scalar{X}} \phi$ and superconformal invariance is retained in the theory below $M$.

Activating \abbr{SUSY} breaking shifts the \abbr{VEV}s of the superfields $X$ and $S$ by $\mathcal{O}(\Fphi)$.  Since $\Fphi \ll M$, the effect on $\vev{\scalar{X}}$ is not important; however, $\vev{\scalar{S}} \sim \Fphi$ once \abbr{SUSY} is broken.  This can be understood from the fact that $\vev{X} = \vev{\scalar{X}} \phi$ implies $\vev{F_X}/\vev{\scalar{X}} = \Fphi$ which, combined with the $F$-term expression \eq{Toy.F.term}, implies $\vev{\scalar{S}} \sim \Fphi$.  Because $\Fphi \ne 0$ breaks superconformal invariance, the induced \abbr{VEV} of $S$ need not preserve the \abbr{AMSB} form.  To check this, one may examine the scalar potential before eliminating the auxiliary fields,
\begin{equation}
\begin{aligned}[b]
V	&=	- \half \abs{F_X}^2 - \half \abs{F_S}^2
		- 2 \lambda_X \scalar{X} \scalar{S} F_{X} + 2 M^2 \Fphi \scalar{S}
		- \inp{ \lambda_X \scalar{X}^2 - M^2 } F_S
		+ \text{h.c.}
\end{aligned}
\label{Eq:Toy.Potential.with.F}
\end{equation}
Minimizing with respect to $\scalar{X}$ gives
\begin{align}
\fderiv{V}{\scalar{X}}
	= - 2 \lambda_X F_{X} \scalar{S} - 2 \lambda_X \scalar{X} F_S = 0
\label{Eq:Toy.X.min}
\end{align}
which can be solved for the ratio of the $F$-term to scalar field \abbr{VEV}s:
\begin{equation}
\frac{\vev{F_S}}{\vev{\scalar{S}}} = - \frac{\vev{F_X}}{\vev{\scalar{X}}} = - \Fphi.
\label{Eq:Toy.VEV.AuxScalarRatio.S}
\end{equation}
The last equal sign of \eq{Toy.VEV.AuxScalarRatio.S} follows from the $X$ threshold preserving \abbr{AMSB}.  Explicit minimization of the potential after substitution of the $F$-terms produces the same result.

From \eq{Toy.VEV.AuxScalarRatio.S} it is seen that the threshold set by $\vev{S}$ is \emph{not} \abbr{AMSB} preserving; rather,
\begin{equation}
\vev{S} = \frac{\vev{\scalar{S}}}{\phi} = - \frac{\Fphi^\dagger}{2 \lambda_X \phi}.
\label{Eq:Toy.VEV.superfield.S}
\end{equation}
This deflection may be communicated to the low-scale theory by introducing a vector-like pair of fields $Y$ and $\Ybar$ (charged under some unspecified gauge group) with the additional term\footnote{As this document was in preparation, \cite{Cai:2010tj} appeared with a similar superpotential to the full superpotential, $\superf{W}_0 + \superf{W}_{\text{Mess}}$, proposed here.} $\superf{W}_{\text{Mess}} = \lambda_Y S Y \Ybar$.  Note that these fields are massless if uncharged under the $R$ symmetry described above.  Below $M$ the $Y$'s acquire a mass
\begin{equation}
\superf{W}^-_{\text{Mess}} = \frac{\lambda_Y \vev{\scalar{S}}}{\phi} Y \Ybar = - \frac{\lambda_Y}{2 \lambda_X} \frac{\Fphi^\dagger}{\phi} Y \Ybar
\label{Eq:Toy.SuperW.below.M}
\end{equation}
leading to deflection.  Recall that both the singlets $X$ and $S$ are heavy, so light singlet deflection does not occur here; rather, the \abbr{AMSB} preserving threshold $M$ induces a superconformal violating mass term for the light $Y$ fields.  These fields then act as messengers of the deflection at a scale naturally around $\Fphi$, but possibly much higher due to the ratio $\lambda_Y/\lambda_X$.

Interestingly, the mass term \eq{Toy.SuperW.below.M} is the precise form of the \abbr{EAM} scenario proposed in \cite{Nelson:2002sa}, and the results of the \abbr{EAM} literature\cite{Nelson:2002sa,Hsieh:2006ig} can be applied directly to the low-scale theory here.  As the original theory had no trace of \abbr{EAM} terms, the heavy singlet deflection scenario 
is an ultraviolet completion of \abbr{EAM}; however, it is important to emphasize that \abbr{EAM} considers deflection occurring near $F_\phi$ where this scenario permits thresholds far above that scale.  This is important because the scale of \abbr{SUSY} breaking in \abbr{AMSB} is $F_\phi$ and it is thus reasonable to expect that physics at this scale would deflect from the \abbr{AMSB} trajectory.  Much above $\Fphi$, however, we expect the rules of Pomarol and Rattazzi to be respected: namely no deflection without a remnant light field.  This is not what happens for heavy singlet deflection---in fact, this deflection is related to physics far above the $F_\phi$ scale and is not associated with a light remnant field.

\subsection{Deflection}
\label{results}

We now explore this deflection in the \abbr{MSSM}.  To retain gauge coupling unification, we take the messengers to be in the $5$ and $\bar 5$ representation of $SU(5)$.  The effective messenger coupling in \eq{Toy.SuperW.below.M} is
\begin{equation}
	\lambda \equiv \frac{\lambda_Y}{2 \lambda_X},
\end{equation} 
which sets the messenger scale, $\Mmess = \lambda F_\phi$, and as a ratio of two dimenionsless couplings can actually be quite large.  The strictest constraint is ensuring that the $X$--$S$ sector is tachyon free, which requires $\Mmess \lesssim M$; otherwise, $\Mmess$ may be arbitrarily high.

Now a general threshold can be parameterized as
\begin{equation}
	\mathcal{M}_{\text{mess}} = \Mmess \left(1 + F_\phi \theta^2 + r F_\phi\theta^2 \right),
\end{equation}
where the $\Mmess\left(1+F_\phi \theta^2\right) = \Mmess \phi$ is the \abbr{AMSB} conserving part of the threshold and $r$ measures deflection.  Assuming no visible sector-messenger yukawa couplings, the deflected gaugino and sfermion masses can be calculated at this threshold as in \cite{Hsieh:2006ig}:
\begin{align}
\left. M_{G} \right|_{\Mmess}
	& = \left. M_{G}^{\text{AMSB}} \right|_{\Mmess} + r \Fphi \left. \frac{ \Delta \beta_{g_G} }{ g_G} \right|_{\Mmess}
   \\
\left. m^2 \right|_{\Mmess}
	& = \left. m^2_{\text{AMSB}} \right|_{\Mmess} + \fourth r \inp{r + 2} \abs{\Fphi}^2 \left. \pderiv{\gamma}{g_G} \Delta \beta_{g_G} \right|_{\Mmess}
\end{align}
where $\gamma$ is the anomalous dimension of the scalar (identical above and below the threshold), $\Delta \beta_{g_G}$ is the difference between beta function above and below the threshold, and the \abbr{AMSB} expressions are given in Eqs.~\eqn{AMSB.scalar.masses} and \eqn{AMSB.gaugino.masses}.

From equation \eq{Toy.SuperW.below.M} we see that we have $\mathcal{M}_{\text{mess}} = \Mmess \left(1 - F_\phi \theta^2 \right)$ so that $r = -2$. This means
\begin{align}
\left. M_G \right|_{\Mmess}
	& = \left. M_{G}^\text{AMSB} \right|_{\Mmess} - 2\Fphi \left. \frac{ \Delta \beta_{g_G} }{ g_G} \right|_{\Mmess}
	\\
\left. m^2 \right|_{\Mmess}
	& = \left. m^2_\text{AMSB} \right|_{\Mmess},
\end{align}
indicating that scalar masses will not be directly affected by the deflection but rather are altered through \abbr{RGE} effects of the gaugino masses.  It is therefore possible to lift the slepton masses with a large number of messengers ($N_5$) or a large messenger scale (set by $\lambda$).  We therefore explore this scenario for different values of $\lambda$ versus the number of messengers.

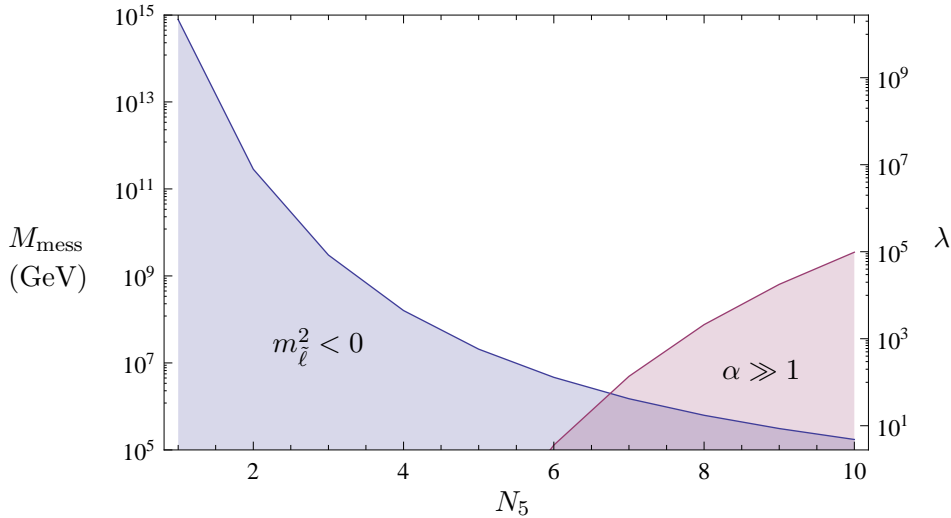
\begin{figure}
\begin{center}
		\begin{picture}(288,180)
			\put(0,0){\includegraphics[scale=1]{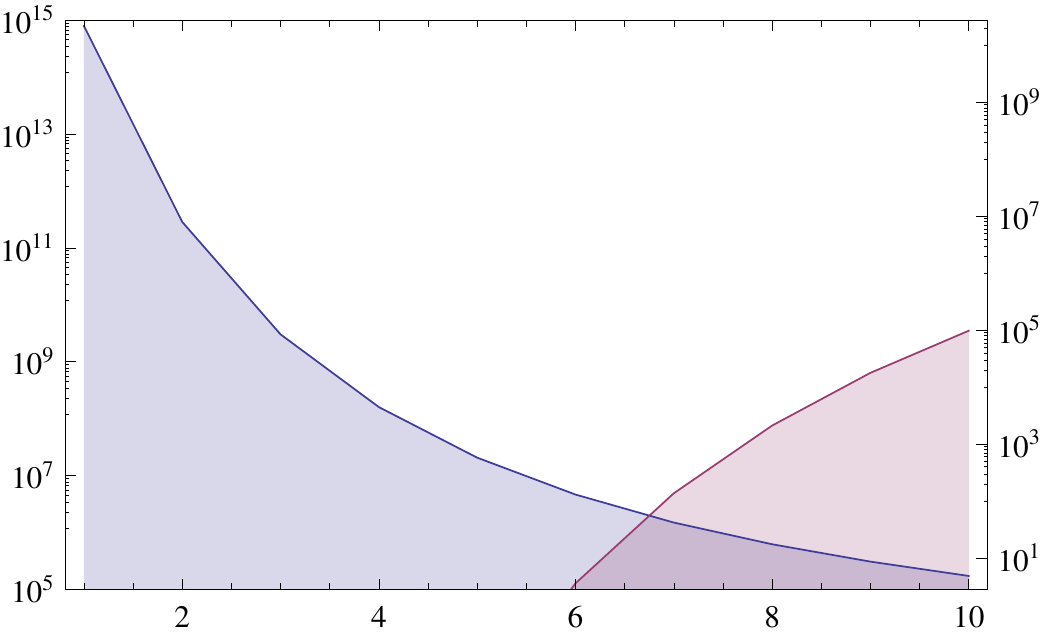}}
			\put(310, 90){$\lambda$}
			  \put(230, 40){$\alpha \gg 1$}
			\put(-40, 90){$\Mmess$}
			\put(-40, 75){(GeV)}
			\put(144, -10){$N_{5}$}
			  \put(60, 50){$m_{\tilde{\ell}}^2 < 0$}
		\end{picture}
\end{center}
	\caption
	{
		$\Mmess$ and $\lambda$ versus the 
		number of messengers.  The light purple region labeled $m_{\tilde{\ell}}^2 < 0$ and the red region denoted $\alpha \gg 1$ are excluded due to tachyonic sleptons and non-
		perturbativity of the gauge coupling up to the \abbr{GUT} scale, respectively.}
	\label{Fig:S.Light.Values}
\end{figure}

It is sufficient to check that the right-handed slepton masses are larger than current bounds to ensure that all slepton masses have been lifted. The allowed parameter space is shown in \fig{S.Light.Values}.  Here the light purple region labeled $m_{\tilde{\ell}}^2 < 0$ is excluded due to tachyonic sleptons whilst the red region labeled $\alpha \gg 1$ is excluded because of non-perturbativity of the gauge coupling up to the \abbr{GUT} scale.  
 
The spectrum has the general form
\begin{equation}
 	m_{\tilde g} > m_{\tilde q} > m_{\tilde W} > m_{\tilde B} >m_{l_L}>m_{l_R}
\end{equation}
with about an order of magnitude difference between the colored fields and right-handed sleptons.  The \abbr{LSP} is the right-handed stau and therefore \abbr{LHC} signals will be dominated by charged tracks and leptonic signals such as described in \cite{DeSimone:2009ws}.  The model as-is lacks a dark matter candidate; however, the strong $CP$ problem may be solved using a Peccei-Quinn symmetry\cite{Peccei:1977hh,Peccei:1977ur,Weinberg:1977ma,Wilczek:1977pj} which due to \abbr{SUSY} yields an axino in the spectrum.  Even if $R$-parity is conserved, the stau can decay into the axino with a lifetime of the order of seconds, and is therefore cosmologically innocuous\cite{Covi:2009pq}.  There are alternative possibilities such as the inclusion of $R$-parity violating operators\cite{Covi:2009pq,Steffen:2008qp}, or adding an extra singlet, as in \cite{Hsieh:2006ig,Cai:2010tj}, which could also resolve this issue.

\section{Conclusion}

We have shown that non-supersymmetric thresholds, which are a necessary ingredient for deflecting off of the \abbr{AMSB} trajectory, are possible even in cases where the threshold is not controlled by a light field (as it is in \cite{Pomarol:1999ie}).  The key ingredient is the stabilization of the potential via non-supersymmetric terms, which break the conformal invariance of \abbr{AMSB}.  If the corresponding \abbr{VEV} also serves as a threshold for some messengers, then the conformal breaking is broadcast to the low-scale theory resulting in deflection.  This can lead to deflection in scenarios previously thought to exhibit decoupling behavior---remarkably even possessing high-scale deflecting thresholds---and indicates that the class of thresholds which lead to \abbr{AMSB} deflection is larger than commonly believed.  
The simplest example of this mechanism is a \abbr{UV}-completion of the \abbr{EAM} scenario which requires non-WIMP dark matter and sleptons significantly lighter than the rest of the SUSY particles.

\section{Acknowledgements}

We are indebted to Z.~Chacko and Y.~Shirman for many useful discussions.  We would also like to acknowledge P.~Fileviez Perez, M.~Luty, R.~N.~Mohapatra, and R.~Rattazzi for providing comments.  This work was supported in part by the National Science Foundation grant no.~Phy-0354401.  N.S.~is also supported by the Australian Research Council.  S.S.~is additionally supported by the U.S.~Department of Energy under grant No.~DE-FG02-95ER40896 and the Wisconsin Alumni Research Foundation.

\bibliography{%
\bibpath/amsb_uv_sensitivity%
}

\end{document}